\documentclass[aps,prl,twocolumn,groupedaddress]{revtex4}

\usepackage{amsbsy,amssymb,amsmath,bm}
\usepackage{graphicx,color}

\begin{document}

\preprint{APS/123-QED}

\title{Asymmetric quintuplet condensation in the frustrated $S$=1 spin dimer compound Ba$_3$Mn$_2$O$_8$}

\author{E. C. Samulon$^1$, Y. Kohama$^{2, 3}$, R. D. McDonald$^2$, M. C. Shapiro$^1$, K. A. Al-Hassanieh$^4$, C. D. Batista$^4$, M. Jaime$^2$, I. R. Fisher$^1$}

\affiliation{$^1$Geballe Laboratory for Advanced Materials and Department of Applied Physics, Stanford University, Stanford, California 94305, USA}

\affiliation{$^2$National High Magnetic Field Laboratory, Los Alamos National Laboratory, Los Alamos, New Mexico 87545, USA}

\affiliation{$^3$Materials and Structures Laboratory, Tokyo Institute of Technology, Yokohama, 226-8503, Japan}

\affiliation{$^4$Theoretical Division, Los Alamos National Laboratory, Los Alamos, New Mexico 87545, USA}

\begin{abstract}

Ba$_3$Mn$_2$O$_8$ is a spin-dimer compound based on pairs of $S=1$, 3$d^2$, Mn$^{5+}$ ions arranged on a triangular lattice.  Antiferromagnetic intradimer exchange leads to a singlet ground state in zero field, with excited triplet and quintuplet states at higher energy.  High field thermodynamic measurements are used to establish the phase diagram, revealing a substantial asymmetry of the quintuplet condensate.  This striking effect, all but absent for the triplet condensate, is due to a fundamental asymmetry in quantum fluctuations of the paramagnetic phases near the various critical fields.

\end{abstract}

\pacs{75.40.-s, 74.20.Mn, 75.30.-m, 75.45.+j}

\maketitle

Spin dimer compounds provide fertile ground for studying novel ordered states in the context of a magnetic material \cite{Rice_2002}.  Typically comprising $S$=1/2 ions, antiferromagnetic intradimer exchange leads to a groundstate that is a product of singlets.  Interaction between dimers causes the excited triplet states to broaden in energy, while application of an external magnetic field leads to familiar Zeeman splitting. Above a characteristic critical field the minimum in the $S_z=1$ triplon band crosses the singlet in energy, leading to long range order at T=0 \cite{Sachdev_2004, Tachiki_1970}.  Depending on the balance of potential and kinetic energy of the triplets, the ground state can be variously approximated as a Bose-Einstein condensate (canted XY antiferromagnet) \cite{BEC, Nikuni_2000, Ruegg_2003, Jaime_2004}, a triplet crystal (inhomogeneous Ising order) \cite{Kodama_2002} or even a supersolid \cite{Sengupta_2007}.  Significantly, quantum fluctuations play a prominent role in these systems, providing means to dynamically tune quantities such as the quasiparticle effective mass \cite{Zapf_2006}.

\begin{figure}
\includegraphics[width=8cm]{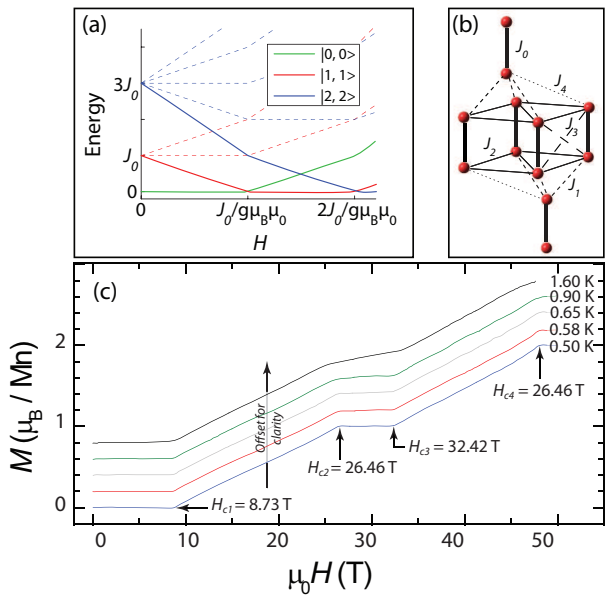}
\caption{(Color online) (a) Energy spectrum versus field for isolated dimer composed of two spin 1 ions. (b) Schematic diagram illustrating the coordination of Mn dimers in Ba$_3$Mn$_2$O$_8$.  $J_0$-$J_4$ represent exchange interactions, described in main text.  (c) Magnetization curves for Ba$_3$Mn$_2$O$_8$ taken for fields applied perpendicular to the $c$ axis. Successive temperature sweeps are offset by 0.2 $\mu_B$ for clarity.}
\label{Mag}
\end{figure}

An intriguing facet of spin dimer compounds, which has not been widely studied, is the possibility of higher moment ordered states at fields beyond that required to saturate the triplet states.  Considering the example of dimers composed of spin 1 ions, the energy spectrum consists in zero magnetic field of excited triplet and quintuplet states above the singlet ground state (Fig. \ref{Mag}(a)).  Application of a magnetic field splits both the triplet and quintuplet states, and depending on the size of interactions, can lead at low temperatures to a distinct $S_z=1$ triplet/ $S_z=2$ quintuplet ordered state at fields above those required to the saturate the triplet states.  The organic biradical magnet F2PNNNO, composed of a strongly ferromagnetically linked spin 1/2 pair linked antiferromagnetically to another pair, acts as a spin 1 dimer \cite{Hosokoshi_1999, Tsujii_2007}.  Experiments have shown two regions of linearly increasing magnetization, first as the $S_z=1$ triplet states and later as the $S_z=2$ quintuplet states are populated, in addition to an ordered phase for the first region. In this letter we explore the nature of the quintuplet condensate in the novel spin dimer compound Ba$_3$Mn$_2$O$_8$, finding a substantial asymmetry that can be accounted for by consideration of quantum fluctuations near the critical fields.

Ba$_3$Mn$_2$O$_8$ crystallizes in the rhombohedral R$\bar{3}$m structure, and is comprised of pairs of Mn$^{5+}$ 3$d^2$ $S=1$ ions arranged vertically on hexagonal layers (Fig. \ref{Mag}(b)) \cite{Weller_1999}.  Inelastic neutron scattering (INS) measurements revealed the dominant intradimer exchange ($J_0=1.642$ meV), as well as a hierarchy of additional exchange interactions, including the dominant exchanges linking dimers within a hexagonal plane ($J_2-J_3=0.1136$ meV) and the dominant out of plane interactions ($J_1=0.118$ meV and $J_4=0.037$) \cite{Stone_2008a, Stone_2008b}.  Electron paramagnetic resonance (EPR) experiments in the diluted compound Ba$_3$(V$_{1-x}$Mn$_x$)$_2$O$_8$ (where the V$^{5+}$ 3$d^0$ ion carries no moment) reveal an almost isotropic $g$-tensor, with $g_{cc}=1.96$ and $g_{aa}=1.97$, and a single ion uniaxial anisotropy characterized by $D$ = 5.81 GHz, corresponding to 0.024 meV \cite{Whitmore_1993}. Similar measurements for the pure compound Ba$_3$Mn$_2$O$_8$ indicate a zero field splitting of the triplet characterized by $|D|$ = 0.032 meV \cite{Hill_2007}, the difference being due to the presence of additional symmetric anisotropies in the concentrated lattice, in particular dipolar coupling between the two ions on each dimer.  Thermodynamic measurements in fields up to 30T revealed a complex phase diagram associated with the triplet states, consisting of two distinct ordered states for fields oriented away from the $c$ axis, attributed to the competition between single ion anisotropy and interdimer exchange in the frustrated lattice \cite{Samulon_2008}. Here we probe the behavior of single crystals of Ba$_3$Mn$_2$O$_8$ via magnetization, heat capacity and magnetocaloric effect measurements in fields large enough to close the gap to the $S_z=2$ quintuplet states.

Extraction magnetization measurements were performed in pulsed magnetic fields up to 60T in a $^3$He refrigerator for fields applied perpendicular to the $c$ axis \cite{Detwiler_2000}. Data were obtained by integrating the field derivative of the magnetization and are plotted for increasing fields in Figure \ref{Mag}(c).  The data were cross calibrated with low field SQUID measurements to attain absolute values of the magnetization.  Critical fields, evident as discontinuities in the slope of the magnetization, were determined from peaks in the second derivative of magnetization with respect to field.  At base temperature, the data show no magnetization up to $H_{c1}$=8.73T, followed by linearly increasing magnetization as the dimers are populated with triplet states, and a plateau at 1 $\mu_B$ between $H_{c2}$=26.46T and $H_{c3}$=32.42T corresponding to one $S_z=1$ triplet per dimer.  There is then a second region of linearly increasing magnetization as the quintuplet band is filled, before finally reaching the full saturation magnetization of 2 $\mu_B$ at $H_{c4}$=47.9T \cite{Uchida_2002}.

\begin{figure}
\includegraphics[width=8cm]{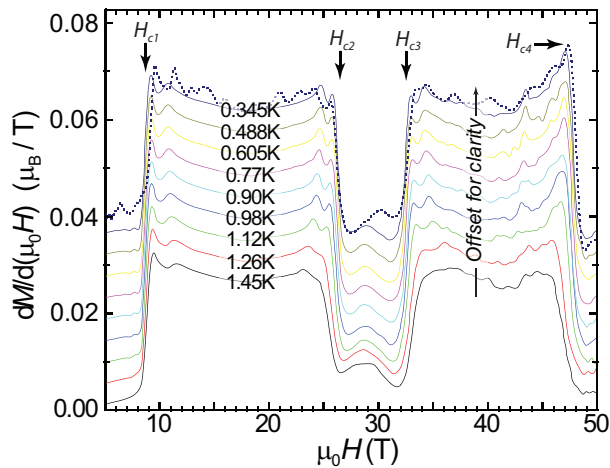}
\caption{(Color online) Field derivative of magnetization of Ba$_3$Mn$_2$O$_8$ for increasing fields applied perpendicular to the $c$ axis.  Successive temperature sweeps offset by 0.0045 $\mu_B$/T for clarity.  Dotted line shows representative magnetization derivative for decreasing fields at 0.345K, revealing no significant hysteresis.}
\label{dMdB}
\end{figure}

The derivative of magnetization with respect to field is plotted in Figure \ref{dMdB} for various temperatures. The data were obtained by measuring the time derivative of both the magnetization and field, and taking the quotient. The signal to noise ratio diminishes at high fields, for which the field derivative tends to 0.  These curves, shown for increasing fields, reveal clear peaks associated with each of the critical fields.  A representative curve for decreasing field at base temperature is plotted revealing the same qualitative behavior.  For falling fields the field derivative is smaller, resulting in a worse signal to noise ratio.  The data also show a second peak to the right (left) of $H_{c1}$ and $H_{c3}$ ($H_{c2}$ and $H_{c4}$).  These peaks do not mark phase transitions but rather signify a zero temperature crossover between two distinct phases observed in our earlier work \cite{Samulon_2008}.

Inspection of Fig. \ref{dMdB} reveals that at lowest temperature the peak at $H_{c4}$ is significantly higher than the peaks at the other three critical fields.  The magnitude of the peaks at the critical fields is set by the number of low-energy excited states available at each critical point. At each critical field $H_{c\nu}$ (where $\nu=\{1,2,3,4\}$), the system can be described as a dilute gas of gapless bosons with a dispersion relation $\omega^{\nu}(\bf q)$ that is quadratic at low energies. Since each boson is a quasiparticle that carries a unit of magnetization ($S^z=1$) along the field direction, the boson density is the magnetization per site $\langle S^z_{j} \rangle$.  The diagonal components of the  mass tensor are $m^{-1}_{\nu i}= \partial^2\omega^{\nu}(\mathbf q)/\partial^2 q_i|_{\mathbf {q}=\mathbf{Q}}$, where $i=\{x,y,z\}$ and ${\mathbf Q}$ is the wavevector where $\omega^{\nu}(\mathbf q)$ is minimized.  The enhanced peak at $H_{c4}$ implies a larger density of states, and by extension a larger effective mass for the corresponding bosons. The difference in the effective mass is a direct result of the strong asymmetry in the zero point fluctuations of the paramagnetic phases. Conservation of total $S^z$ implies that such fluctuations are absent above $H_{c4}$. In contrast, zero-point fluctuations consisting of creation (annihilation) of $S^z=2$ quintuplet-singlet pairs (triplet-triplet pairs) are present for $H_{c2} \leq H \leq H_{c3}$. Their effect is to reduce the mass of the bosons. This can be easily appreciated by a thought experiment in which the field is fixed at some arbitrary value between $H_{c2}$ and $H_{c3}$ and the interdimer exchange couplings are increased by applying pressure. The applied pressure $P$ will close the gap at a critical point ($P=P_c$) where $H_{c2}$ becomes equal to $H_{c3}$, causing the single-particle excitation spectrum to become linear ($z=1$). Such an $XY$ quantum phase transition in $D=d+1=4$ dimensions is driven by the zero-point (phase) fluctuations under consideration \cite{UpperExponent}. A simple continuity argument implies that the boson mass has to be reduced by such fluctuations in order to get a massless spectrum when they diverge at $P=P_c$.

In the dilute limit relevant for the regions around each of the quantum critical points, the inter-particle distance $\rho^{-1/3}$ (where $\rho$ is the boson density) is much larger than  the scattering length $a$ (lattice parameter) of the hard-core repulsion between bosons. The effective boson-boson repulsion, $v_0 = \Gamma_{\bf 0}({\bf Q},{\bf Q})$, is obtained from a ladder summation \cite{Beliaev_1958}:
\begin{equation}
\Gamma_{\bf q} \left(\bf k, \bf {k'} \right) = V_{\bf q} - \int_{-\pi}^{\pi} \frac{d^3p}{8\pi^3} V_{\bf q - \bf p} \frac{\Gamma_{\bf p} \left(\bf k, \bf {k'} \right) }{\omega^{\nu}({\bf k}+{\bf p})+\omega^{\nu}({\bf k'}-{\bf p})}.
\end{equation}
Here $V_{\bf q}=U + \left(J_2 + J_3\right) \gamma^2_{\bf q} + \frac{J_1}{2} \gamma^1_{\bf q} + \frac{J_4}{2} \gamma^3_{\bf q}$, where $U \rightarrow \infty$ comes from the hard-core repulsion, the rest of the terms come from nearest neighbor repulsion terms and $\gamma^n_{\bf q}$'s are defined as follows:
\begin{eqnarray}
\gamma^1_{\bf q} & = & \mathrm{cos} \left(q_3\right) + \mathrm{cos} \left( q_3 - q_1 \right) + \mathrm{cos} \left( q_3 - q_2 \right)
\nonumber\\
\gamma^2_{\bf q} & = & \mathrm{cos} \left(q_1\right) + \mathrm{cos} \left(q_2\right) + \mathrm{cos} \left( q_1 - q_2 \right)
\nonumber\\
\gamma^3_{\bf q} & = & \mathrm{cos} \left(q_3 + q_1 - q_2 \right) + \mathrm{cos} \left(q_3 - q_1 + q_2 \right)
\nonumber\\
&& + \mathrm{cos} \left(q_3 - q_1 - q_2 \right),
\end{eqnarray}
where $q_n$'s refer to the principal reciprocal axes.  The dispersions $\omega^{\nu}({\bf q})$ were computed by using a generalized spin-wave approach \cite{Khaled}. For the paramagnetic (PM) region near each QCP, a mean field treatment of the effective interaction $v_0$ leads to a renormalized chemical potential ${\tilde \mu_{\nu}}= \mu_{\nu} - 2 v_0 \rho$ with $\mu_{\nu}=g \mu_B \left(-1\right)^{\nu} \left(H_{c\nu}-H\right)$ \cite{Nikuni_2000}. Low temperature thermodynamic properties can be computed from the renormalized chemical potential in the PM region around each QCP. The striking asymmetry between $H_{c4}$ and the rest of the critical fields follows from the fact that the corresponding masses $m_{4\parallel}=1.9357K^{-1}$ and $m_{4\perp}=0.4531K^{-1}$ are roughly two times bigger than the masses at the other three critical fields: $m_{3\parallel}=0.7834K^{-1}$, $m_{3\perp}=0.1833K^{-1}$, $m_{2\parallel}=0.6977K^{-1}$, $m_{2\perp}=0.1633K^{-1}$, and $m_{1\parallel}=0.9133K^{-1}$, $m_{1\perp}=0.2138K^{-1}$.

In general, the mass asymmetry present in Ba$_3$Mn$_2$O$_8$ induces asymmetries in all of the thermodynamic properties. For instance, the condensation temperature (${\tilde \mu}(T=T_c)=0$) near the critical fields is: $T_c \propto \rho^{2/3} m^{-1}_{\nu} \propto |H-H_{c\nu}|^{2/3} m^{-1/3}_{\nu},$ where we have used that $\Gamma^{\nu}_{\bf 0} ({\bf Q},{\bf Q}) \propto m^{-1}_{\nu}$. The increased effective mass at $H_{c4}$ reduces the ordering temperature, causing the phase boundary to be steeper near $H_{c3}$ than near $H_{c4}$. In general, all spin dimer compounds should exhibit some degree of asymmetry because zero-point fluctuations are absent only in the highest field paramagnetic phase.  The effect is especially striking in Ba$_3$Mn$_2$O$_8$ because $H_{c3}-H_{c2} \sim 6$T is considerably smaller than  $H_{c4}-H_{c3} \sim15.5$T, implying the gap in energy to create a zero-point fluctuation near $H_{c3}$ is relatively small.

\begin{figure}
\includegraphics[width=7cm]{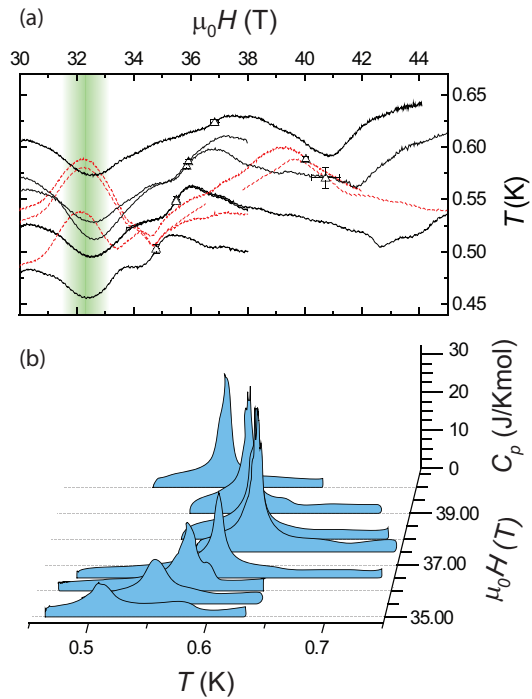}
\caption{(Color online)  (a) MCE curves plotted versus field for increasing field (solid black lines) and decreasing fields (dashed red lines) for fields applied perpendicular to the $c$ axis.  Phase transitions can be determined from peaks in derivative with respect to field, and are shown as open triangles.  Vertical shading indicates broad maximum associated with rapidly changing temperature derivative of magnetization at the critical field. (b) Heat capacity, taken on cooling, as a function of temperature for fields applied perpendicular to the $c$ axis.}
\label{Cp}
\end{figure}

To observe the additional consequences of the mass asymmetry, we measured the high field phase diagram by both magnetocaloric effect (MCE) and heat capacity experiments performed in a 45T hybrid superconducting/resistive magnet at $^3$He temperatures for fields perpendicular to the $c$ axis.  For MCE measurements, transitions are marked by a sharp increase (decrease) in temperature on entering (leaving) the ordered phase, while in practice the phase boundary is determined from peaks in the derivative (see inset to Fig. \ref{Cp}(a)).  The vertical shaded region in Fig. \ref{Cp}(a) refers to a temperature independent effect, associated with a rapidly changing temperature derivative of magnetization at $H_{c3}$ \cite{Samulon_2008}.

\begin{figure}
\includegraphics[width=8.5cm]{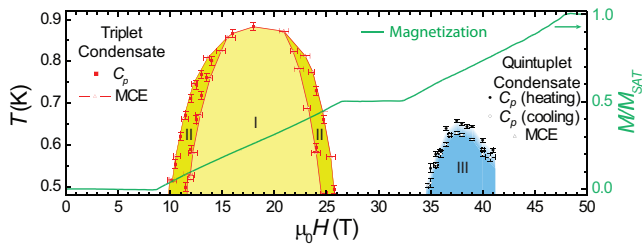}
\caption{(Color online) Phase diagram for fields applied perpendicular to the $c$ axis.  Representative magnetization curve for $T$ = 0.50 K is shown for reference (right axis).  Blue region corresponds to the quintuplet condensate.  Solid and open circles correspond to heat capacity data taken on heating and cooling respectively, illustrating the apparent hysteresis caused by the experimental technique.  Dark and light yellow regions correspond to phase II and phase I of the triplet condensate as defined in ref. \cite{Samulon_2008}.}
\label{Phase}
\end{figure}

Heat capacity measurements were performed using standard thermal relaxation time calorimetry, in the large $\Delta T$ limit. Accordingly, $C_p$ is calculated from the instantaneous time derivative of the calorimeter temperature and the conductance of the thermal link as discussed elsewhere \cite{Lashley_2003}.  For these experiments, the primary benefit of this technique is its expediency, which unfortunately comes at the cost of reduced accuracy.  The reduced accuracy caused a small extraneous hysteresis on heating versus cooling. To improve the accuracy, multiple temperature curves were measured at each field and then averaged.  The data taken on cooling are plotted in Fig. \ref{Cp}.  They reveal an increasing critical temperature for fields leading up to 37T which has the maximum $T_c$ of 0.63K.  Above 37T the critical temperature begins to decrease again.  $T_c$ values extracted from heat capacity data for both cooling and heating are shown as a lower and upper bound on the phase boundary in Fig. \ref{Phase}.  The integrated entropy under the curve appears to track the critical temperature, so that a higher $T_c$ implies more entropy. This effect is also produced by the asymmetry in the zero-point fluctuations that shift the maximum of the order parameter towards the critical field with larger fluctuations ($H_{c3}$ for the triplet-quintuplet phase) \cite{Khaled}.

The entire phase diagram for fields applied perpendicular to the $c$ axis is shown in Figure \ref{Phase} together with a magnetization curve, taken at 0.50K, showing the four critical fields as discontinuities in the slope.  All told the phase diagram reveals at least three distinct ordered states across a large field range.  Inspection of this figure reveals a striking asymmetry in the quintuplet condensate, for which the maximum $T_c$ of the quinton phase is found at $H$=37T, less than 5T from $H_{c3}$ but more than 10T from $H_{c4}$.  Additionally, a small degree of asymmetry was found in the ordered states of the singlet-triplet regime, with phase II slightly narrower on the high field side than the low field side. Both asymmetries are expected from the different masses reported above: $m_{2i}/m_{1i} \simeq 0.7639 $ and $m_{4i}/m_{3i} \simeq 2.4711$. The measured ratio between maximum $T_c$'s, $T^{st}_{max}/T^{tq}_{max} \sim $ 1.37, agrees well with the ratio of $4/3$ between the effective interdimer exchange interactions obtained from the two level model \cite{Samulon_2008} for the singlet-triplet and triplet-quintuplet phases \cite{Khaled}.

In summary, we have determined the high field phase diagram of the $S=1$ dimer compound Ba$_3$Mn$_2$O$_8$.  Quantum fluctuations play a profound role in determining the relative symmetry/asymmetry of the triplet/quintuplet condensates, respectively.  This material provides an ideal test bed for studying and tuning the effect of zero-point fluctuations.

The authors acknowledge experimental assistance from N. Harrison and V. Zapf. Work at Stanford University is supported by the National Science Foundation, Division of Materials Research under grant DMR-0705087. Crystal growth equipment purchased with support from the Department of Energy, Office of Basic Energy Sciences, under contract DE-AC02-76SF00515. Most of the experimental portion of this work was performed at the NHMFL, which is supported by NSF Cooperative Agreement No. DMR-0084173, by the State of Florida, and by the DOE.  The theoretical portion of this work was carried out under the auspices of the NNSA of the U.S. DOE at LANL under Contract No. DE-AC52-06NA25396.


\begin{thebibliography}{1}
\bibitem{BEC} Small anisotropic terms will always break the continuous symmetry of spin rotations along the field axis required for a Bose-Einstein condensate.  However, if those terms are much smaller than the dominant interdimer exchange then the ground state can be well approximated by a BEC.
\bibitem{UpperExponent} The gapless mode that arises at such a pressure induced quantum critical point is linear in momentum.  This leads to the effective dimension of the QCP as D = 3+1 (d=3 and z=1).
\end{thebibliography}
\end{document}